\documentclass[onecolumn,usenatbib,usegraphicx]{mn2e}
\usepackage{amsmath}
\usepackage{amssymb}

\def\be{\begin{equation}}
\def\ee{\end{equation}}
\def\Msun{\,M$_{\odot}\,$}
\def\vph{$v_{\rm ph}$}
\def\vphm{v_{\rm ph}}

\begin{document}

\title[Radiation-Driven Instabilities in Massive Stars]{Local Radiation-Driven Instabilities in Post-Main Sequence Massive Stars}

\author[Suarez-Madrigal, A., Krumholz, M.~R. \& Ramirez-Ruiz, E.]
{Andr\'es Su\'arez-Madrigal$^1$, Mark R. Krumholz$^1$\thanks{E-mails: krumholz@ucolick.org (MRK), enrico@ucolick.org (ER)}, \& Enrico Ramirez-Ruiz$^1$\\
Department of Astronomy and Astrophysics, University of California Santa Cruz\\ 211 Interdisciplinary Sciences Building, 1156 High Street, Santa Cruz, CA 95064, USA}

\maketitle
\begin{abstract}
Late in their evolution, massive stars may undergo periods of violent instability and mass loss, but the mechanism responsible  for these episodes has not been identified. We study one potential contributor: the development of local radiation-driven instabilities in the outer layers of main sequence (MS) and post-MS massive stars. We construct a sequence of massive stellar evolution models and investigate where they are subject to local radiative instabilities, both in the presence of magnetic fields and without them, and at a range of metallicities. We find that these types of instabilities do not occur in solar-metallicity MS stars up to 100 $M_\odot$, but they set in immediately post-MS for stars heavier than $\sim 25$ $M_\odot$. Once an instability appears, it involves a significant amount of mass in the star's upper layers  (up to $\sim 1$\,per cent of the initial stellar mass), suggesting that radiation-driven instabilities are a potentially viable  mechanism for dynamic mass loss. We find that the presence of magnetic fields at strengths low enough not to disturb the hydrostatic balance of the star does not alter these results. Stars with sub-Solar metallicity also show instability, but their instabilities involve less mass and appear later in the star's evolution. 
\end{abstract}


\section{Introduction}

The evolution of massive stars is driven largely by mass loss \citep[e.g.][]{chiosi86a}. This loss takes the form of both continuously driven winds \citep[and references therein]{puls08a} and episodes of violent mass ejection, although which dominates the overall mass loss budget is still disputed (e.g.~\citealt{smith06a} versus \citealt{vink12a}). While the continuous winds can be understood as the product of resonant scattering of photons off metal atoms in the stellar atmosphere \citep{castor75b, vink00a, vink01a}, the origin of the eruptive episodes is much less certain.
Supernova observations, in particular of Type IIn, show evidence  for massive stars undergoing sizable  eruptive outbursts ($\gtrsim 10^{-3}M_\odot$) within months to years prior to their explosion \citep[e.g][]{dopita84, chugai94, smith07, foley07,pastorello12, ofek13}.

The prototypical example of violent mass loss is the Great Eruption of $\eta$ Carinae in the 1840s \citep{davidson97a}, which ejected several tens of $M_\odot$ over a 20 year period \citep{gomez06a, gomez10a}. The rate of mass loss rules out the line scattering that is responsible for the continuous winds as a possible mechanism \citep{owocki04a}. Instead, proposed models fall into two broad classes. In one, the mass loss is powered by the internal radiation of the star interacting with a source of continuum rather than line opacity, and is thought to be triggered by the development of clumps in the stars' upper layers that causes a rapid change in the effective continuum opacity \citep[e.g.][]{shaviv00a, shaviv01b, smith06a, van-marle08a}. The other class of models invokes tidal perturbation by a companion as the driving mechanism. The perturbation, applied to a star that is already on the edge of stability, induces an episode of mass transfer that gives rise to an outburst and mass loss \citep[e.g.][]{soker04a, soker05a, kashi10a}.

Either of the proposed mechanisms calls for the star to be on the verge of instability, so that some perturbation, either internal or external, can trigger an outburst. Since the amount of mass involved is large, this instability cannot be confined to the stellar atmosphere but must involve deeper layers of the star. There are several candidate global instabilities that affect stars near the Eddington limit, including strange modes \citep{glatzel99a}, global radiatively-driven modes \citep{shaviv01a}, and opacity-driven $g$-modes \citep{townsend06a}.

However, thus far there has been no significant search for local instabilities that might play a role in destabilizing the outer layers of massive stars, making them ripe for eruptive ejection. This is somewhat surprising, because the criterion for local radiation-hydrodynamic and radiation-magnetohydrodynamic instabilities in optically thick, gravity-confined media was derived very generally by \citet{blaes03a}. Indeed, \citeauthor{blaes03a} conjecture that the outer layers of massive stars might be subject to their computed instability \citep{fernandez12}.

In this paper we investigate that conjecture, and determine where, in terms of both stellar mass and evolutionary state, local radiation-driven instabilities in stars will occur. We construct a grid of evolutionary tracks for the interiors of stars at a variety of masses using the \verb!MESA! (Modules for Experiments in Stellar Astrophysics) package \citep{paxton11a}, and determine where in that grid stars satisfy the \citet{blaes03a} instability condition. We show that local instability does not occur in main sequence stars, but that it appears almost immediately after the end of hydrogen burning in stars with initial masses above $\sim 25$ $M_\odot$. When present, the instability occurs at high optical depth within the star, and the unstable region typically contains up to 1 $M_\odot$ of material, suggesting that this instability is a candidate contributor  to eruptive mass loss.

The remainder of this paper is structured  as follows. In Section \ref{sec:methods} we present the details of our models and stability calculations. We report the results of this exercise in Section \ref{sec:results}, and discuss their implications in Section \ref{sec:discussion}.

\section{Methodology}
\label{sec:methods}
In this section, we first describe the instability criteria we will use, and then discuss how we generate the stellar structure models to which we will apply them.


\subsection{Radiation-driven hydro- and magneto-hydrodynamical instabilities}

Starting from the equations of radiation hydrodynamics in the diffusion limit, \cite{blaes03a} obtain dispersion relations for local radiation pressure-driven 
instabilities in the local Wentzel--Kramers--Brillouin (WKB) limit. Their analysis relies on a few assumptions: in the unperturbed state (but not necessarily in the perturbed state) matter and radiation are in LTE, the matter is of constant composition and ionization state, gravitational potential perturbations are negligible (the Cowling approximation), as is photon viscosity. They consider both purely hydrodynamic (HD) and magneto-hydrodynamic (MHD) cases; for the latter, the magnetic field in the unperturbed state is required to be constant.

For the HD case, the dispersion relation for the acoustic waves ($\omega \propto k$), to first order, is given by equation (62) of \cite{blaes03a} and reproduced here:
\be
        \omega = \pm k c_i - i \frac{\kappa_F}{2 c c_i}\left(1+\frac{3P_{\rm gas}}{4E}\right)\left[\left(\frac{4E}{3}+P_{\rm gas}\right)c_i \mp (\mathbf{\hat{k} \cdot F})\Theta_{\rho}\right],
        \label{eq:dispHydro}
\ee
where $\mathbf{\hat{k}}$ is the wave vector, $c_i$ is the isothermal sound speed in the gas, $\kappa_F$ is the flux mean opacity, $c$ is the speed of light, $P_{\rm gas}$ is the gas pressure, $E$ and ${\bf F}$ are, respectively, the radiation energy density and flux, and $\Theta_{\rho}$ is the logarithmic derivative of $\kappa_F$ with respect to the density, i.e.
\be
\Theta_{\rho} \equiv \frac{\partial \ln{\kappa_F}}{\partial \ln{\rho}}.
\ee
From this dispersion relation, the condition for instability is
\be
        \eta \equiv - \frac{\kappa_F}{2 c c_i}\left(1+\frac{3P_{\rm gas}}{4E}\right)\left[\left(\frac{4E}{3}+P_{\rm gas}\right)c_i \mp (\mathbf{\hat{k} \cdot F})\Theta_{\rho}\right] > 0.
        \label{eq:condHydro}
\ee
For $\eta>0$, $\eta$ is the growth rate of the instability.
Since all physical quantities involved are positive, instability occurs only when the second term inside square brackets is larger than the first term and opposite in sign, that is, for the upward propagating acoustic waves corresponding to the minus in the plus-minus sign. The first term represents damping by radiative diffusion, while the second represents forcing by the radiation flux.
Note that the forcing term is proportional to the logarithmic derivative of the opacity $\Theta_\rho$. This will become significant in our analysis below.

The dispersion relation for magneto-acoustic waves, to first order, is given by equation (92) of \cite{blaes03a} and reproduced here:
\begin{align}
        \omega = \pm k \vphm - i \frac{\kappa_F}{2 c \vphm}\left[\frac{\vphm^2-(\mathbf{\hat{k} \cdot v_A})^2}{2\vphm^2-v_A^2-c_i^2}\right]\left(1+\frac{3P_{\rm gas}}{4E}\right)\left[\left(\frac{4E}{3}+P_{\rm gas}\right)\vphm \mp (\mathbf{\hat{k} \cdot F})\Theta_{\rho}\right] \notag \\
        \pm \frac{i \kappa_F}{2c\vphm(2\vphm^2-v_A^2-c_i^2)}\left(1+\frac{3P_{\rm gas}}{4E}\right)(\mathbf{\hat{k} \cdot v_A})(\mathbf{\hat{k} \times v_A}) \cdot (\mathbf{\hat{k} \times F}),
        \label{eq:dispMHD}
\end{align}
where $v_A$ is the magnitude of the vector Alfven speed ($\mathbf{v_A} \equiv \mathbf{B}/[4 \pi \rho]^{1/2}$), and \vph\ is the phase speed of the wave, defined as
\be
\vphm \equiv \frac{1}{2}\{v_A^2+c_i^2 \pm [(v_A^2+c_i^2)^2 - 4(\mathbf{\hat{k} \cdot v_A})^2 c_i^2]^{1/2}.
\ee
This dispersion relation gives the condition for instabilities to occur, i.e.
\begin{align}
        \eta \equiv - \frac{\kappa_F}{2 c \vphm}\left[\frac{\vphm^2-(
\mathbf{\hat{k} \cdot v_A})^2}{2\vphm^2-v_A^2-c_i^2}\right]\left(1+\frac{3P_{\rm gas}}{4E}\right)\left[\left(\frac{4E}{3}+P_{\rm gas}\right)\vphm \mp (\mathbf{\hat{k} \cdot F})\Theta_{\rho}\right] \notag \\
        \pm \frac{\kappa_F}{2c\vphm(2\vphm^2-v_A^2-c_i^2)}\left(1+\frac{3P_{\rm gas}}{4E}\right)(\mathbf{\hat{k} \cdot v_A})(\mathbf{\hat{k} \times v_A}) \cdot (\mathbf{\hat{k} \times F}) > 0
        \label{eq:condMHD};
\end{align}
again, when instability occurs, $\eta$ gives the growth rate.
We note that in this case there are two terms involved in the condition: the first one is a magnetically-modified hydrodynamic term, analogous to the one presented in equation (\ref{eq:condHydro}) and in which the opacity driving must overcome the radiative diffusion to contribute to an instability.
However, the inclusion of a magnetic field in the analysis introduces the factor
\be
\left[\frac{\vphm^2-(\mathbf{\hat{k} \cdot v_A})^2}{2\vphm^2-v_A^2-c_i^2}\right],
\label{eq:magnetmod}
\ee
which is always positive.
The second term in equation (\ref{eq:condMHD}) corresponds to a purely magnetohydrodynamic contribution, and is positive both for ``fast'' and ``slow'' magnetosonic waves. As such, this term always favors instability, but will be important only when it dominates the hydrodynamic term.


\subsection{Stellar models in MESA star}
\label{sec:mesamodels}

We will use the instability conditions given by equations (\ref{eq:condHydro}) and (\ref{eq:condMHD}) on stellar profiles of stars computed using \verb!MESA star! to determine whether instabilities might be able to develop in stars at a range of masses and ages. \verb!MESA star! is a one-dimensional stellar evolution code included in the \verb!MESA! package for computational stellar astrophysics \citep{paxton11a}. \verb!MESA star! is a Henyey style code that simultaneously solves the fully coupled structure and composition equations; we refer readers to \citet{paxton11a} for details on its capabilities and limitations.

We use \verb!MESA star! to simulate 21 stars with different masses, all with Solar metalicity, using the default code parameters. Our model stars have masses from 5\Msun to 100 M$_\odot$, in intervals of 5 M$_\odot$, plus a model for a 1\Msun star. This range in masses allows us to study differences in behavior between low mass and high mass stars. In order to understand how metallicity affects the results, we also perform a similar set of runs at metallicities of 1, 3, 10, and 20\,per cent of Solar metallicity. At each output time for each mass, we evaluate the instability conditions given by equations (\ref{eq:condHydro}) and (\ref{eq:condMHD}) at every position within the structure of the calculated stellar profiles.
We take the partial derivatives of the opacity $\Theta_\rho$ directly from the \verb!Mesa star! output.

There are two complications to our analysis. First, \verb!MESA star! does not include magnetic fields. To evaluate the MHD instability condition, we therefore adopt constant magnetic fields at a range of strengths. Second, \citet{blaes03a} only derive their instability criteria for conditions of constant mean molecular mass, and this assumption breaks down within stars' hydrogen and helium ionization zones. For this reason, we exclude from our analysis all those locations where the hydrogen ionization fraction is $1-99$\,per cent, or where the helium ionization fraction is $10-90$\,per cent. This ensures that we are only considering regions where the mean molecular mass is constant to 1\,per cent or better. Note that this exclusion does not mean that there cannot be instabilities present in the ionization zones, simply that the instability criteria to which we have access are not applicable in those regions.

\section{Results}
\label{sec:results}
\subsection{Hydrodynamic Instabilities in Single Stars}

\begin{figure}
\begin{center}
        \includegraphics[scale=0.45,angle=0]{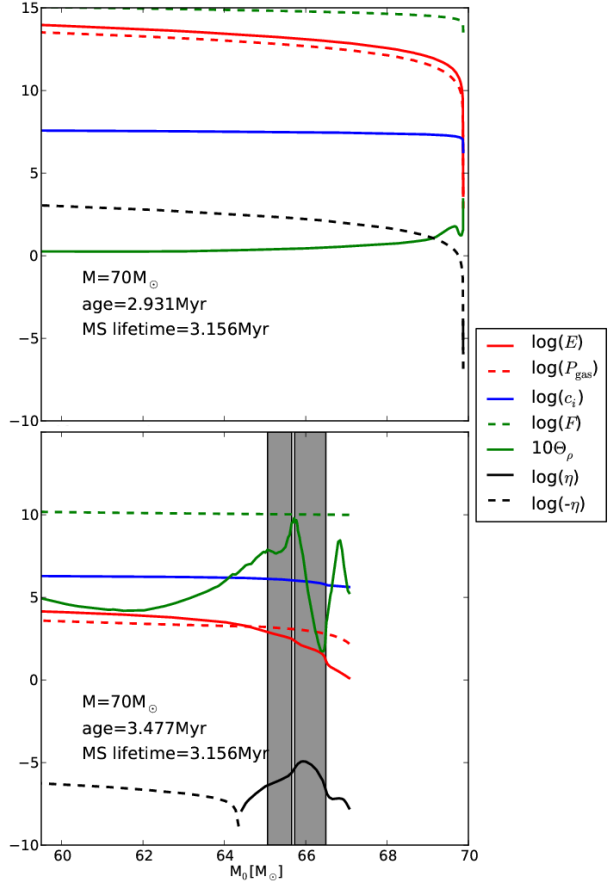}
\end{center}
        \caption{Properties of a star with an initial mass of  70\Msun as a function of the mass coordinate in two stages of its evolution. All dimensional quantities are measured in cgs units. The upper panel shows the state of the star at an age of 2.931 Myr, while the star is still on the main sequence, while the lower panel shows its state at 3.477 Myr, after the end of hydrogen burning at 3.156 Myr. Note that the mass of the star is $<70$\Msun due to stellar winds. The quantities plotted are those required to evaluate the HD growth rate $\eta$ (equation \ref{eq:condHydro}), including $\eta$ itself. Dashed black lines indicate regions where $\eta < 0$ and the star is stable, while solid black lines indicate regions where $\eta > 0$ and the star is unstable. The gray shaded bars mark the hydrogen and helium ionization zones that we exclude from our analysis; see Section \ref{sec:mesamodels}. The upper panel shows a star without instabilities, while the lower one presents a stage in which the star has already developed an opacity-driven instability.
        }
        \label{fig:70msunHydroprops}
\end{figure}

We first consider the purely HD case at Solar metallicity; we defer discussion of magnetic and metallicity effects to Section~\ref{ss:z_and_b}. Figure \ref{fig:70msunHydroprops} shows sample results from one of our models: we plot the physical properties of a 70\Msun star as a function of the mass coordinate at two different moments of its life. The quantities shown are those required to evaluate the HD instability condition (equation~\ref{eq:condHydro}). The upper panel shows the state of the star while it is on the main sequence; as the plot shows, at this time the system does not yet show any relevant instability. As discussed above, the condition for an instability to be present is that the radiative flux term must overcome diffusive damping; the former is proportional to the logarithmic derivative of the opacity with density $\Theta_\rho$, and during the main sequence stage $\Theta_\rho \sim 0.01$ throughout most of the star.  The underlying physical reason for this small value is that, on the main sequence, the stellar material is very hot even fairly far out in the star. As a result, the opacity is dominated by electron scattering which is density-independent, rather than free-free or bound-free processes which are not. As a result, radiative forcing is weak and the star is stable.

The lower panel, on the other hand, shows the star after it leaves the main sequence. At this point an acoustic instability has developed, even excluding the ionization zones where we cannot apply the \citet{blaes03a} instability analysis.  Instability occurs in two zones near the surface of the star, on either side of the hydrogen and helium ionization zones. There is a notable difference between the unstable zones: in the innermost, the radiation pressure $E$ and the gas pressure $P_{\rm gas}$ are at the same order of magnitude, while in the region that reaches the stellar surface $P_{\rm gas}$ is much greater than $E$. The total mass contained in the two unstable zones together is $\sim 1$\Msun at this evolutionary stage (Figure \ref{fig:70msunHydroprops}). The underlying reason for the instability is that, as the star leaves the main sequence, it expands and the gas temperature in the outer layers drops drastically; this causes free-free and bound-free opacity to become competitive with electron scattering, which in turn raises $\Theta_\rho$ and destabilizes the gas.

\subsection{Hydrodynamic Instabilities as a Function of Stellar Mass and Age}

\begin{figure}
\begin{center}
        \includegraphics[scale=0.45,angle=0]{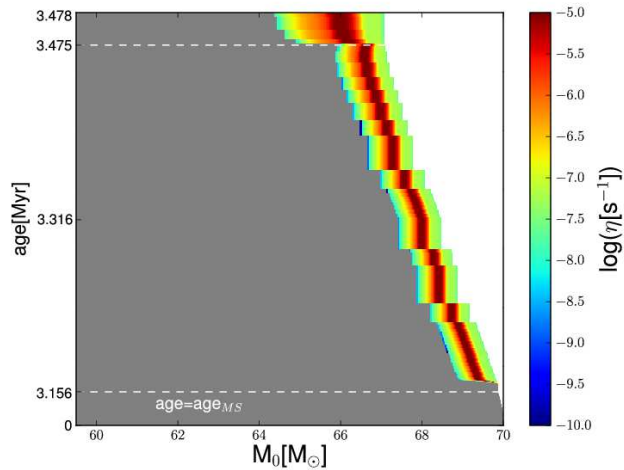}
 \end{center}
        \caption{Growth rate ($\eta$) for HD instabilities in the outer regions of a 70\Msun star, as a function of its age. Gray regions indicate where $\eta < 0$, so the star does not satisfy the HD instability condition. Color regions indicate values of $\eta > 0$. Note that the vertical axis has three different linear scales for the age (separated by white dashed lines) in order to show the behavior of different evolutionary stages of the star: its MS (up to 3.156Myr), a post-MS stage in which the star gradually loses mass and builds a layer of unstable material (from 3.156Myr to 3.475Myr) and a later stage in which the star becomes more unstable towards its surface (from 3.475Myr to 3.478Myr).
        \label{fig:70msunHydromap}
        }
\end{figure}

By repeating the analysis above for a given star at different stages of its evolution, we can find where and when the instability appears for a given initial mass. Figure \ref{fig:70msunHydromap} shows the growth rate of the HD instability computed over the outer 15\,per cent of the mass of our initial 70\Msun star as a function of its age. Negative values for $\eta$, indicating stability, are shown in grey. As the star loses mass, a white region appears on the right side of the figure, representing this decrement. The end of the main sequence for the star is shown with the lower white dashed line and a label. To accommodate in a single graph the varying timescales present during the stellar evolution, the age axis is broken in three different linear scales, separated by the two white dashed lines in the image. This figure shows that the HD instability condition (equation~\ref{eq:condHydro}) is fulfilled in the outermost region of the star almost immediately after it finishes core hydrogen burning. The amount of mass contained in the unstable region tends to grow as the star evolves. The inverse of the growth rate is a characteristic time of the instability; for most of the unstable zone, the growth time is smaller than $\sim 10^8$\,s (i.e. $\eta \gtrsim 10^{-8}$\,s$^{-1}$), which amounts to a few years and is significantly less than the age of the star. 
We also find that the wavelength of the most unstable mode is much smaller than either the stellar radius or the radial extent of the unstable region, indicating that instabilities will not be quenched by the finite thickness of the unstable layer.

\begin{figure}
\begin{center}
        \includegraphics[scale=0.4,angle=0]{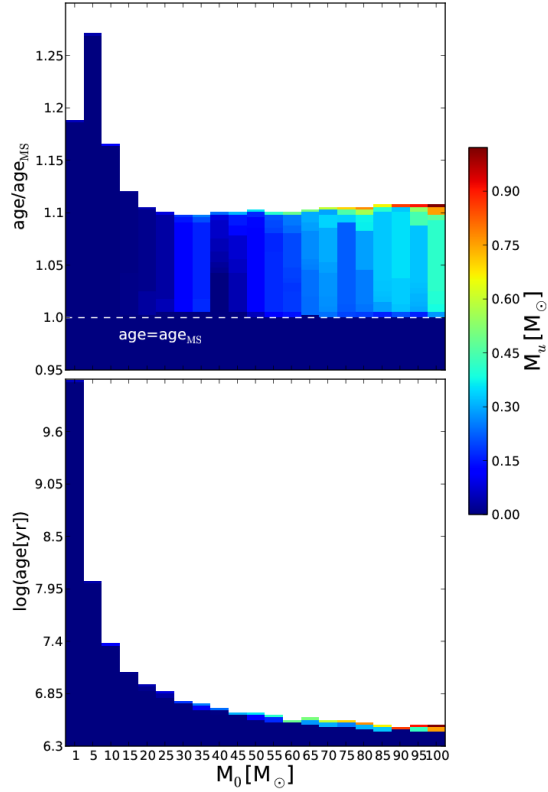}
\end{center}
        \caption{Mass contained in the unstable regions ($M_u$) of stars as a function of their initial mass (horizontal axis) and age (vertical axis). Upper panel: the age of each star is shown in units of its MS lifetime. Lower panel: ages are shown in years, with the highest value plotted corresponding to the current age of the Universe.
        \label{fig:mapaMinestablesHydro}
        }
\end{figure}

By extending our analysis to stars of different masses, we can compute the mass $M_u$ in the unstable region for each star as a function of initial mass $M_0$ and stellar age. Note that our calculation of $M_u$ excludes the ionization zones, so it  provides a lower limit for the size of the unstable region. Figure \ref{fig:mapaMinestablesHydro} shows our results. For both panels, the horizontal axis shows initial mass, while the vertical axis represents the age of each star. In the upper panel we measure age in units of each model's MS lifetime, defined as the time at which hydrogen burning in the core ceases; the white dashed horizontal line indicates this moment (age = age$_{\rm MS}$). The lower panel shows the same result, but now with the vertical axis indicating absolute rather than normalized age. The upper limit of the vertical axis in this plot is $\approx 14$\,Gyr, comparable to the age of the Universe; we therefore see that a 1\,\Msun star will not become unstable over the entire age of the Universe. From these maps, it is clear that although most of the stars reach a stage where they have appropriate conditions for acoustic radiative-driven instabilities, it is only massive stars (above $\sim 25$\Msun) that actually have a significant fraction of their mass contained in the unstable region, and where the instability appears rapidly after the end of the main sequence rather than extremely late in their lives.

\begin{figure}
\begin{center}
        \includegraphics[scale=0.4,angle=0]{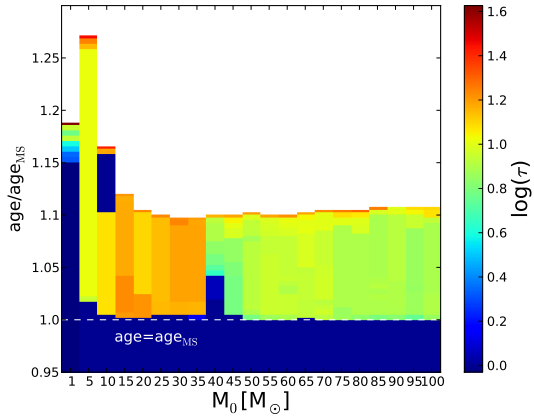}
 \end{center}
        \caption{Colors show the Rosseland mean optical depth ($\tau$) from the stellar surface to the inner edge of the outermost unstable region.
Axes are identical to those in the upper panel of Figure \ref{fig:mapaMinestablesHydro}. Unstable regions extend significantly deeper than the stellar atmosphere (defined at $\tau = 2/3$).
        \label{fig:mapatauHydro}
        }
\end{figure}

It is also of interest to investigate how deep within the star the unstable regions lie. In Figure \ref{fig:mapatauHydro} we plot the Rosseland mean 
optical depth $\tau$ from the stellar surface to the inner edge of the outermost unstable region.
The coordinate axes in the figure are the same as those in the upper panel of Figure \ref{fig:mapaMinestablesHydro}: age in units of MS age versus initial stellar mass. This graph demonstrates that the unstable regions of interest have a high optical depth, which suggests that this mechanism can produce a layer of unstable mass lying deeper than the stellar photosphere at $\tau =2/3$. It is interesting to note that even the models in which the fraction of mass in the outer unstable region is minuscule, the optical depth is quite high. When instabilities are present, they typically go down to $\tau \sim 10$.

\subsection{Effects of Metallicity and Magnetic Fields}
\label{ss:z_and_b}

\begin{figure}
\begin{center}
        \includegraphics[scale=0.4,angle=0]{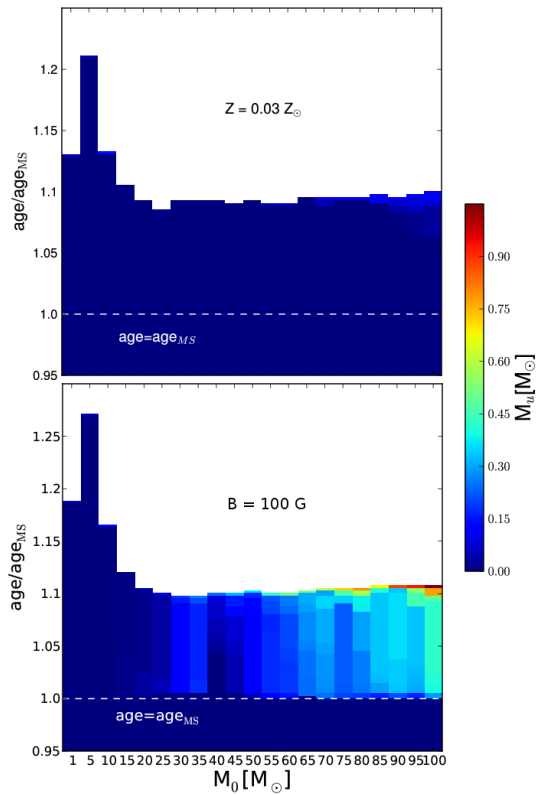}
 \end{center}
        \caption{Color maps showing the amount of mass contained in the outermost unstable regions of each model at different stages of their evolution. The upper panel shows models for which the metallicity has been decreased to 0.03 $Z_\odot$: the amount of unstable mass involved has decreased in all stars, but the general tendency seen before remains.
The models depicted in the lower panel are identical to the ones in the upper panel in Fig. \ref{fig:mapaMinestablesHydro}, but a constant-magnitude magnetic field of 100\,G has been included. The effect in the unstable masses is indistinguishable.
        \label{fig:mapa_metal_MHD}
        }
\end{figure}

All the stellar models we have considered so far have Solar metallicity: $Z_\odot$. Here we extend our formalism  to include sub-Solar metallicity models with  0.2, 0.1, 0.03, and 0.01 $Z_\odot$. We find that the qualitative result that the mass in the unstable region increases with initial stellar mass, while the time post-MS at which instability appears decreases, continues to hold for varying metallicity. 
However, the amount of mass in the unstable region (again excluding the ionization zones) decreases for all models as the metallicity decreases. We illustrate this effect in the upper panel in Figure \ref{fig:mapa_metal_MHD}, which shows a map similar to that plotted in the upper panel of Figure \ref{fig:mapaMinestablesHydro}, but for 0.03 $Z_\odot$. The general weakening of the instability at lower metallicity can again be understood by considering the density-dependence of the dominant opacities, and thus the value of $\Theta_\rho$. As the metallicity decreases, the supply of heavy ions in the outer layers of the star decreases, and thus the importance of bound-free opacity is reduced compared to the electron scattering opacity. Since the former is density-dependent and the latter is not, the overall opacity becomes less density-dependent ($\Theta_\rho$ is reduced) and the radiative force is diminished. This weakens the instability.

We also investigated the MHD instability condition (equation~\ref{eq:condMHD}) using constant-magnitude magnetic fields ranging from 1\,G to 10$^6$\,G.
In order to maintain the static equilibrium presumed by \cite{blaes03a}, and to remain within the range of validity of the non-magnetic \verb!MESA star! models, we only consider magnetic fields small enough that the magnetic pressure is significantly smaller than the gas pressure. A 10$^6$\,G magnetic field amounts to a pressure of $\approx$ 4$\times 10^{10}$\,Ba, significantly lower than the gas and radiation pressures present during the MS of a 70\Msun star, as seen in the upper panel of Figure \ref{fig:70msunHydroprops}; on the other hand, the strongest magnetic field that we could safely propose for the same star when the instabilities are present is of the order of 100\,G, corresponding to a magnetic pressure of $\approx$ 400\,Ba (see lower panel of Figure \ref{fig:70msunHydroprops}).
 This restriction is reasonable, since there is no evidence that magnetic fields in stellar interiors are ever strong enough for magnetic pressure to significantly alter their hydrostatic balance, and it seems unlikely that one could construct a hydrostatic spherical stellar model if they were. With this restriction in mind, we find that magnetic fields in the allowed range of strengths for already developed instabilities have negligible effects on the size of the instability region. The hydrodynamical contribution to the instability always dominates. The lower panel in Figure \ref{fig:mapa_metal_MHD} presents the case where a magnetic field of magnitude $B=100$\,G was set up. This map is practically indistinguishable from its HD counterpart (upper panel in Fig. \ref{fig:mapaMinestablesHydro}) and illustrates the lack of a magnetic effect in this situation. Applying stronger fields (up to $10^6$\,G) to a 70\Msun star during its MS does not contribute to the formation of instabilities.

We can understand these results by examining the form of the instability condition. Late in the star's evolution, when radiative forcing dominates the hydrodynamic term in equation (\ref{eq:condMHD}), the ratio between the hydrodynamic and magnetic terms is of order $(v_{\rm ph}/v_A)^2 \Theta_\rho$. If the magnetic field is dynamically weak compared to gas pressure, then $v_{\rm ph} \approx c_i$ and the term becomes of order $(c_i/v_A)^2 \Theta_\rho$. Thus the magnetic contribution to the instability condition is significant only if $\Theta_\rho \ll (v_A/c_i)^2$. 
Figure \ref{fig:70msunHydroprops} shows that $\Theta_\rho$ is typically of order a few tenths in the outer parts of post-MS stars. In this post-MS case, this means that there is no room for magnetic fields to have a significant effect. The requirement that magnetic pressure be small compared to gas pressure implies that $(v_A/c_i)^2 \ll 1$, but magnetic fields only significantly alter the instability condition if $(v_A/c_i)^2 \gg \Theta_\rho \sim 0.1$. This analysis shows that it is clearly impossible for the magnetic field to have anything more than a marginal effect. 
In the MS case, where the HD term is dominated by radiative diffusion damping, the ratio of HD and magnetic terms is of order $(c_i^2/v_A^2) c_i(E+P_{\rm gas})/F$, assuming again that the magnetic field is weak compared to the radiation and gas pressures. 
In this case we can make a similar argument. The radiative flux $F$ is of order $cE/\tau$, where $\tau$ is the characteristic optical depth, which means that the factor $c_i(E+P_{\rm gas})/F \sim \tau (P_{\rm gas}/E) (c_i/c)$. The factor $c_i/c \sim 10^{-3}$ is small, but deep in the stellar interior $\tau \gg 10^3$, and far out in the star $P_{\rm gas}/E \gg 10^3$, so the total factor is everywhere $\gg 1$; examination of Figure \ref{fig:70msunHydroprops} indicates that it is of order $10^4 - 10^6$. This large value is a direct consequence of the fact that stellar interiors are in the dynamic rather than the static diffusion limit \citep{krumholz07b}. At the same time, the restriction that the magnetic field not dominate the hydrostatic balance of the star implies that $(c_i^2/v_A^2)  \gtrsim 1$. Together these two constraints imply that the HD term must greatly exceed the magnetic one in main sequence stars.
In practice and in theory, we find that including magnetic fields has little practical effect in the behavior of the instabilities already present in the purely HD case.

\section{Conclusions and implications}
\label{sec:discussion}

In this paper we investigate under what conditions local radiation-driven instabilities will be present in stellar interiors. To this end we construct a series of stellar models as a function of mass and age using the \verb!MESA star! code \citep{paxton11a}, and then apply the hydrodynamic and magnetohydrodynamic instability criteria derived by \citet{blaes03a}. We find that that the outer layers of massive stars can present the properties required for instability to develop. This does not occur in main sequence stars up to $100$ M$_\odot$, but it sets in immediately post-MS for stars more massive than $\sim25$ M$_\odot$. Once these instabilities develop, they involve a non-negligible amount of the stellar mass, reaching up to $\sim1$\,\Msun in the more massive cases studied. They also occur relatively deep in the stellar interior, at optical depths of order $\tau = 10$.
The underlying physical cause for the onset of the instability is a drop in gas temperature as stars expand in the post-MS phase, which raises the importance of bound-free and free-free opacity compared to electron scattering. Bound-free and free-free opacity are density-dependent, and this density-dependence is destabilizing.

This process could be a potential mechanism for the enhanced dynamic mass loss that is observed in some massive stars during their lifetimes and prior to exploding.
The presence of a layer of material that is unstable in this way could prepare the conditions of an episode of eruptive mass loss, perhaps triggered by some external event such as the gravitational interaction of the star with a potential binary companion. At a minimum, the presence of the instability will modify the structure of the star in the regions where it occurs. However, determining the full results of the instability will require detailed numerical simulations.

\section*{Acknowledgments}
It is a pleasure to thank O.~Blaes and A.~Socrates for helpful discussions. AS thanks T.~Sukhbold, M.~Macleod and the \verb!MESA! users list for direction in the appropriate use of \verb!MESA star! for this work, as well as A.~Heger who was kind enough to share a few stellar models for comparison. We acknowledge support from the Alfred P.~Sloan Foundation (AS and MRK), the David and Lucile Packard Foundation (AS and ER), NASA through a Chandra Space Telescope  grant (MRK and ER) and Astrophysics Theory Program grant NNX09AK31G (MRK), and the NSF through grants CAREER-0955300 (MRK) and AST-0847563 (ER).

\bibliographystyle{mn2e}
\bibliography{refs}

\begin{thebibliography}{}

\bibitem[\protect\citeauthoryear{{Blaes} \& {Socrates}}{{Blaes} \&
  {Socrates}}{2003}]{blaes03a}
{Blaes} O.,  {Socrates} A.,  2003, \apj, 596, 509

\bibitem[\protect\citeauthoryear{{Castor}, {Abbott} \& {Klein}}{{Castor}
  et~al.}{1975}]{castor75b}
{Castor} J.~I.,  {Abbott} D.~C.,    {Klein} R.~I.,  1975, \apj, 195, 157

\bibitem[\protect\citeauthoryear{{Chiosi} \& {Maeder}}{{Chiosi} \&
  {Maeder}}{1986}]{chiosi86a}
{Chiosi} C.,  {Maeder} A.,  1986, \araa, 24, 329

\bibitem[\protect\citeauthoryear{{Chugai} \& {Danziger}}{{Chugai} \&
  {Danziger}}{1994}]{chugai94}
{Chugai} N.~N.,  {Danziger} I.~J.,  1994, \mnras, 268, 173

\bibitem[\protect\citeauthoryear{{Davidson} \& {Humphreys}}{{Davidson} \&
  {Humphreys}}{1997}]{davidson97a}
{Davidson} K.,  {Humphreys} R.~M.,  1997, \araa, 35, 1

\bibitem[\protect\citeauthoryear{{Dopita}, {Cohen}, {Schwartz} \&
  {Evans}}{{Dopita} et~al.}{1984}]{dopita84}
{Dopita} M.~A.,  {Cohen} M.,  {Schwartz} R.~D.,    {Evans} R.,  1984, \apjl,
  287, L69

\bibitem[\protect\citeauthoryear{{Fern\'andez} \& {Socrates}}{{Fern\'andez} \&
  {Socrates}}{2012}]{fernandez12}
{Fern\'andez} R.,  {Socrates} A.,  2012, \apj

\bibitem[\protect\citeauthoryear{{Foley}, {Smith}, {Ganeshalingam} \& {et
  al.}}{{Foley} et~al.}{2007}]{foley07}
{Foley} R.~J.,  {Smith} N.,  {Ganeshalingam} M.,    {et al.} 2007, \apj, 657,
  105

\bibitem[\protect\citeauthoryear{{Glatzel}, {Kiriakidis}, {Chernigovskij} \&
  {Fricke}}{{Glatzel} et~al.}{1999}]{glatzel99a}
{Glatzel} W.,  {Kiriakidis} M.,  {Chernigovskij} S.,    {Fricke} K.~J.,  1999,
  \mnras, 303, 116

\bibitem[\protect\citeauthoryear{{Gomez}, {Dunne}, {Eales} \&
  {Edmunds}}{{Gomez} et~al.}{2006}]{gomez06a}
{Gomez} H.~L.,  {Dunne} L.,  {Eales} S.~A.,    {Edmunds} M.~G.,  2006, \mnras,
  372, 1133

\bibitem[\protect\citeauthoryear{{Gomez}, {Vlahakis}, {Stretch}, {Dunne},
  {Eales}, {Beelen}, {Gomez} \& {Edmunds}}{{Gomez} et~al.}{2010}]{gomez10a}
{Gomez} H.~L.,  {Vlahakis} C.,  {Stretch} C.~M.,  {Dunne} L.,  {Eales} S.~A.,
  {Beelen} A.,  {Gomez} E.~L.,    {Edmunds} M.~G.,  2010, \mnras, 401, L48

\bibitem[\protect\citeauthoryear{{Kashi} \& {Soker}}{{Kashi} \&
  {Soker}}{2010}]{kashi10a}
{Kashi} A.,  {Soker} N.,  2010, \apj, 723, 602

\bibitem[\protect\citeauthoryear{{Krumholz}, {Klein}, {McKee} \&
  {Bolstad}}{{Krumholz} et~al.}{2007}]{krumholz07b}
{Krumholz} M.~R.,  {Klein} R.~I.,  {McKee} C.~F.,    {Bolstad} J.,  2007, \apj,
  667, 626

\bibitem[\protect\citeauthoryear{{Ofek}, {Sullivan}, {Cenko} \& {et
  al.}}{{Ofek} et~al.}{2013}]{ofek13}
{Ofek} E.~O.,  {Sullivan} M.,  {Cenko} S.~B.,    {et al.} 2013, \apj

\bibitem[\protect\citeauthoryear{{Owocki}, {Gayley} \& {Shaviv}}{{Owocki}
  et~al.}{2004}]{owocki04a}
{Owocki} S.~P.,  {Gayley} K.~G.,    {Shaviv} N.~J.,  2004, \apj, 616, 525

\bibitem[\protect\citeauthoryear{{Pastorello}, {Cappellaro}, {Inserra} \& {et
  al.}}{{Pastorello} et~al.}{2012}]{pastorello12}
{Pastorello} A.,  {Cappellaro} E.,  {Inserra} C.,    {et al.} 2012,
  arXiv:1210.3568v2

\bibitem[\protect\citeauthoryear{{Paxton}, {Bildsten}, {Dotter}, {Herwig},
  {Lesaffre} \& {Timmes}}{{Paxton} et~al.}{2011}]{paxton11a}
{Paxton} B.,  {Bildsten} L.,  {Dotter} A.,  {Herwig} F.,  {Lesaffre} P.,
  {Timmes} F.,  2011, \apjs, 192, 3

\bibitem[\protect\citeauthoryear{{Puls}, {Vink} \& {Najarro}}{{Puls}
  et~al.}{2008}]{puls08a}
{Puls} J.,  {Vink} J.~S.,    {Najarro} F.,  2008, \aapr, 16, 209

\bibitem[\protect\citeauthoryear{{Shaviv}}{{Shaviv}}{2000}]{shaviv00a}
{Shaviv} N.~J.,  2000, \apjl, 532, L137

\bibitem[\protect\citeauthoryear{{Shaviv}}{{Shaviv}}{2001a}]{shaviv01a}
{Shaviv} N.~J.,  2001a, \apj, 549, 1093

\bibitem[\protect\citeauthoryear{{Shaviv}}{{Shaviv}}{2001b}]{shaviv01b}
{Shaviv} N.~J.,  2001b, \mnras, 326, 126

\bibitem[\protect\citeauthoryear{{Smith}, {Li}, {Foley} \& {et al}.}{{Smith}
  et~al.}{2007}]{smith07}
{Smith} N.,  {Li} W.,  {Foley} R.~J.,    {et al}. 2007, \apj, 666, 1116

\bibitem[\protect\citeauthoryear{{Smith} \& {Owocki}}{{Smith} \&
  {Owocki}}{2006}]{smith06a}
{Smith} N.,  {Owocki} S.~P.,  2006, \apjl, 645, L45

\bibitem[\protect\citeauthoryear{{Soker}}{{Soker}}{2004}]{soker04a}
{Soker} N.,  2004, \apj, 612, 1060

\bibitem[\protect\citeauthoryear{{Soker}}{{Soker}}{2005}]{soker05a}
{Soker} N.,  2005, \apj, 619, 1064

\bibitem[\protect\citeauthoryear{{Townsend} \& {MacDonald}}{{Townsend} \&
  {MacDonald}}{2006}]{townsend06a}
{Townsend} R.~H.~D.,  {MacDonald} J.,  2006, \mnras, 368, L57

\bibitem[\protect\citeauthoryear{{van Marle}, {Owocki} \& {Shaviv}}{{van Marle}
  et~al.}{2008}]{van-marle08a}
{van Marle} A.~J.,  {Owocki} S.~P.,    {Shaviv} N.~J.,  2008, \mnras, 389, 1353

\bibitem[\protect\citeauthoryear{{Vink}, {de Koter} \& {Lamers}}{{Vink}
  et~al.}{2000}]{vink00a}
{Vink} J.~S.,  {de Koter} A.,    {Lamers} H.~J.~G.~L.~M.,  2000, \aap, 362, 295

\bibitem[\protect\citeauthoryear{{Vink}, {de Koter} \& {Lamers}}{{Vink}
  et~al.}{2001}]{vink01a}
{Vink} J.~S.,  {de Koter} A.,    {Lamers} H.~J.~G.~L.~M.,  2001, \aap, 369, 574

\bibitem[\protect\citeauthoryear{{Vink} \& {Gr{\"a}fener}}{{Vink} \&
  {Gr{\"a}fener}}{2012}]{vink12a}
{Vink} J.~S.,  {Gr{\"a}fener} G.,  2012, \apjl, 751, L34

\end{thebibliography}

\end{document}